\documentclass[11pt]{article}
\usepackage[a4paper,margin=1in]{geometry}
\usepackage{graphicx}
\usepackage{multirow} 
\usepackage{subfig}
\usepackage[cmex10]{amsmath}
\graphicspath{{./figures/}} % go find the figures in figures folder
\usepackage{hyperref} % please comment this line if you do not want hyperlink on references

\newtheorem{theorem}{Theorem}[section]
\newtheorem{lemma}[theorem]{Lemma}

\newtheorem{corollary}[theorem]{Corollary}

\newenvironment{proof}[1][Proof]{\begin{trivlist}
\item[\hskip \labelsep {\bfseries #1}]}{\end{trivlist}}
\newenvironment{definition}[1][Definition]{\begin{trivlist}
\item[\hskip \labelsep {\bfseries #1}]}{\end{trivlist}}

\newcommand{\qed}{\nobreak \ifvmode \relax \else
      \ifdim\lastskip<1.5em \hskip-\lastskip
      \hskip1.5em plus0em minus0.5em \fi \nobreak
      \vrule height0.75em width0.5em depth0.25em\fi}

\title{Randomized $\Delta$-Edge-Coloring via Quaternion of Complex Colors\\ (Extended Abstract)}
\author{
	Tony T. Lee \footnote{Tony T. Lee is with Dept. of Information Engineering, The Chinese University of Hong Kong, Hong Kong and Dept. of Electronics Engineering, Shanghai Jiao Tong University, Shanghai, China. Email: ttlee@ie.cuhk.edu.hk }
	\and
	 Yujie Wan \footnote{Yujie Wan is with Dept. of Information Engineering, The Chinese University of Hong Kong, Hong Kong. Email: wyj009@ie.cuhk.edu.hk}
	\and 
	Hao Guan  \footnote{Hao Guan is with Dept. of Information Engineering, The Chinese University of Hong Kong, Hong Kong. Email: hguan@ie.cuhk.edu.hk}
}

\begin{document}
\maketitle 

\begin{abstract}
	This paper explores the application of a new algebraic method of color exchanges to the edge coloring of simple graphs. Vizing's theorem states that the edge coloring of a simple graph $G$ requires either $\Delta$ or $\Delta+1$ colors, where $\Delta$ is the maximum vertex degree of $G$. Holyer proved that it is {\bf NP}-complete to decide whether $G$ is $\Delta$-edge-colorable even for cubic graphs. By introducing the concept of complex colors, we show that the color-exchange operation follows the same multiplication rules as quaternion. An initially $\Delta$-edge-colored graph $G$ allows variable-colored edges, which can be eliminated by color exchanges in a manner similar to variable eliminations in solving systems of linear equations. The problem is solved if all variables are eliminated and a properly $\Delta$-edge-colored graph is reached. For a randomly generated graph $G$, we prove that our algorithm returns a proper $\Delta$-edge-coloring with a probability of at least 1/2 in $O(\Delta|V||E|^5 )$ time if $G$ is $\Delta$-edge-colorable. Otherwise, the algorithm halts in polynomial time and signals the impossibility of a solution, meaning that the chromatic index of $G$ probably equals $\Delta+1$. Animations of the edge-coloring algorithms proposed in this paper are posted at YouTube \url{http://www.youtube.com/watch?v=KMnj4UMYl7k}.
\end{abstract}

\section{Introduction}

The chromatic index $\chi_e (G)$  of a simple graph $G=(V,E)$, with vertex set $V$ and edge set $E$, is the minimum number of colors required to color the edges of the graph such that no adjacent edges have the same color. A theorem proved by Vizing \cite{west2001introduction} states that the chromatic index is either $\Delta$ or $\Delta+1$, where $\Delta$ is the maximum vertex degree of graph $G$. The graph $G$ is said to be Class 1 if $\chi_e(G)=\Delta$; otherwise, it is Class 2. Except for some particular types of graphs, such as bipartite graphs, it is inherently difficult to classify an arbitrary simple graph. In fact, Holyer has proved in \cite{holyer1981np} that it is {\bf NP}-complete to determine the chromatic index of arbitrary simple graph even if $\Delta=3$.

In this paper, we describe a new algebraic method of color exchanges for edge coloring of simple graphs. The original color-exchange method was devised by Alfred Kempe in his endeavor to prove the four-color theorem \cite{kempe1879geographical}. Although his attempt was unsuccessful, his method remains critical to the final proof given by Appel and Haken \cite{appel1989every}. The Kempe chain method was defined on two-colored vertices. An extension of this method to two-colored edges, called alternating paths, constitutes the basis of the proof of Vizing's theorem and Edmonds' matching algorithm \cite{edmonds1987paths}. By introducing the concept of complex colors, we show that the color-exchange operation performed on alternating paths follows the same multiplication rules as quaternion, and edge coloring is a procedure of variable eliminations. 

We consider each edge $e\in E$ as a pair of links; each link is a half-edge. Let $C$ be the set of colors. The coloring of graph $G=(V,E)$ is a function $c:E\rightarrow C\times C$ defined by assigning a color pair, or a complex color, to each $e\in E$, one color assigned for each link. If $|C|=\Delta$, then a color configuration $c$ of $G$ such that all links incident to the same vertex have different colors can be easily obtained. Suppose that the complex color assigned to edge $e$ is $c(e)=(\alpha,\beta),\alpha,\beta \in C$, then the colored edge $e$ is a {\it variable} if $\alpha \neq \beta$; otherwise, $c(e)=(\alpha,\alpha)$ is a {\it constant} for any color $\alpha$. A proper $\Delta$-edge-coloring of graph $G$ can be achieved by eliminating all variables.

Our edge-coloring algorithm starts with an arbitrary color configuration $c$ of $G$, which may contain variable edges. Applying a sequence of well-defined ``moves'' (color exchanges) of variables in a configuration may lead to other new configurations. Variables can be systematically eliminated when they encounter other variables while moving around the graph. For a graph $G$ with $\Delta \geq 3$, the algorithm can be initialized by a configuration of $G$ with a set of $\Delta$ colors $C=\{c_1,...,c_\Delta \}$. First, we eliminate variables that contain color $c_1$, then remove the remaining variables that contain color $c_2$, and so on. The problem is solved if all variables are eliminated and a properly $\Delta$-edge-colored graph is reached; otherwise, the algorithm halts in polynomial time and signals the impossibility of a solution, meaning that the chromatic index of $G$ probably equals $\Delta+1$, which implies that the graph $G$ could be class 2.

Based on random walks on graphs, the average time that a variable hits another variable in a graph has a polynomial bound \cite{aldous2002reversible} \cite{Aleliunas1979random} \cite{lovsz1993random} . For a randomly generated graph $G$, we prove that our algorithm returns a proper $\Delta$-edge-coloring with a probability of at least 1/2 in $O(\Delta|V||E|^5 )$ time if chromatic index $\chi_e(G)=\Delta$. On the other hand, it returns with absolute certainty if $G$ is a class 2 graph. 

In existing literature, the only known exact algorithms are the $O(2^{|V|/2})$ and $O({1.5039}^{|V|} )$ algorithms for 3-edge-coloring proposed by Beigel and Eppstein \cite{eppstein2001improved} \cite{beigel19953}. Vizing's proof implies an $O(|V||E|)$ time algorithm with $\Delta+1$ colors, which was later improved to $O(|E|\sqrt{|V|\log{|V|}})$ by Gabow et al. \cite{Gabowedgecoliring}. Some approximation algorithms with high probability of success were reported in \cite{grable1997nearly} \cite{dubhashi1998near} \cite{aggarwal2003switch}. Grable and Panconesi \cite{grable1997nearly} proposed an edge coloring algorithm using $(1+\epsilon)\Delta$ colors, which operates in $O(\log{}\log{|V|})$ rounds if $\Delta$ is larger than polylog$|V|$ but smaller than any positive power of $|V|$. Dubhashi, Grable, and Panconesi \cite{dubhashi1998near} proposed another $O(\log{|V|}+\log^s{|V|}\log{}\log{|V|})$ time algorithm using $\Delta+\Delta/\log^s{|V|}$ colors if $\Delta=\Omega(\log^k{|V|})$ for some constants $s,k>0$. For general multigraph with $\Delta=\omega(|V|^2)$, Aggarwal et al. \cite{aggarwal2003switch} proposed an algorithm using $\Delta+o(\Delta)$ colors that runs in $O(|V|^2)$ steps. There are also some heuristics reported in \cite{hilgemeier2003fast} that do not provide any performance guarantees. Our randomized algorithm is the first polynomial time algorithm for $\Delta$-edge-coloring.

In principle, each vertex $v\in V$ and each edge $e\in E$ represent a constraint on coloring of edges, and the entire graph $G$ can be considered as a set of simultaneous equations. The variable elimination procedure of edge coloring is similar to the Gaussian elimination of solving systems of linear equations. A comparison between these two procedures is summarized in Table \ref{tab:comp}.
\begin{table}[htpb]
	\centering
	{\footnotesize
		\begin{tabular}{|c || c| c| }
			\hline 
			& System of linear equations & Edge coloring \\
			\hline \hline 
			Operations & arithmetic operations & color exchanges \\ \hline 
			Constraints	& linear equations & vertices and edges \\ \hline 
			Unknowns & variables &  variable-colored edges \\ \hline 
			Algorithms &	variable elimination & variable elimination\\ \hline 
			Solutions	& consistency &	$\Delta$-colorable\\ \hline 
			No solution & inconsistency & infinite loop $(\chi_e(G)=\Delta+1)$\\ \hline 
			Complexity & polynomial time & randomized polynomial time\\ 
			\hline		
		\end{tabular}	
		}
		\caption{Comparison between system of linear equations and edge coloring of simple graphs. }
	\label{tab:comp}
\end{table}

In spite of the similarity between solving linear equations and edge coloring, the main difference is  recognizing the final state. If a system of linear equations has no solution, the inconsistency of the system can be identified by variable eliminations in polynomial time. Eliminating variable edges of class 2 graphs may result in an infinite loop. Recognizing snarks, class 2 cubic graphs, could have been significant in determining the halting state of edge-coloring algorithms. A polynomial time algorithm for identifying snarks will immediately lead to the conclusion that {\bf P=NP}. 

The rest of this paper is organized as follows. In section 2, we define the color function of incidence graphs, and establish the rules of color exchanges and Kempe walks. In section 3, we describe the variable elimination procedure by Kempe walks, and introduce the concept of canonical configurations. Section 4 is devoted to walks on directional paths. In particular, we give a randomized algorithm based on walks on spanning tree with deflection. In addition, the performance of this randomized algorithm and the characterization of snarks are provided in this section. In section 5, we present experimental results and compare our method with existing heuristics. Section 6 provides a conclusion and discussions on future research.

\section{Color-Exchange Operations of Complex Colors}

This section introduces an algebraic method of color exchanges implemented on the edges of a colored simple graph. We first describe the properties of color function defined on incidence graphs, and then establish the rules of color exchanges and Kempe walks. 

\begin{figure}[htpb]
	\centering
	\includegraphics[scale=0.85]{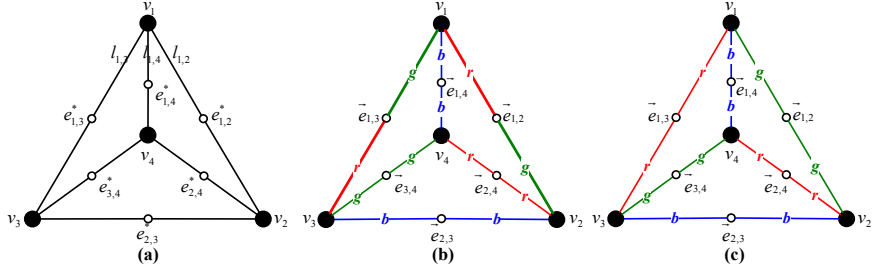}
	\caption{(a) Incidence graph of tetrahedron. (b) Consistent coloring with variables $\vec{e}_{1,3}=(g,r)$ and $\vec{e}_{1,2}=(r,g)$. (c) Proper coloring of tetrahedron.}
	\label{fig:incidencetetrahedron}
\end{figure}

Let $G=(V,E)$ be a simple graph with vertex set $V$, edge set $E$. The {\it incidence graph} $G^*$ is constructed from $G$ by placing a fictitious vertex in the middle of each edge of $G$. Let $E^*(G^*)=\{e_{i,j}^{*}| e_{i,j}\in E(G) \}$ denote the set of fictitious vertices on edges. Then edge $e_{i,j}\in E(G^* )$, connecting two end vertices $v_i$ and $v_j$, consists of two {\it links}, denoted by $l_{i,j}=(v_i,e_{i,j}^{*})$ and $l_{j,i}=(v_j,e_{i,j}^{*})$. Fig. \ref{fig:incidencetetrahedron} (a) illustrates the incidence graph of the tetrahedron.

Let $L(G^*)$ be the set of links and $C=\{c_1,...,c_\Delta \}$ denote a set of $\Delta$ colors. A {\it coloring function} $c$ defined on an incidence graph $G^*$ is a mapping of color assignments on links $c:L(G^* )\rightarrow C$. The color of link $l_{i,j}\in L(G^* )$ is denoted as $c(l_{i,j} )=c_{i,j}$. Since each edge $e\in E(G^* )$ consists of two links, the color function $c$ can also be considered as a mapping defined on the set of edges: $c:E(G^* )\rightarrow C\times C$.	

We define the {\it colored edge} $c(e_{i,j} )= \vec{e}_{i,j}=(c_{i,j},c_{j,i} )$ as a two-tuple color vector, where $c_{i,j}=c(l_{i,j} )$ and $c_{j,i}=c(l_{j,i} )$ are respective colors of the two links of $e_{i,j}$. The following properties of a color function $c$ are related to the edge coloring of graph $G$. 

\begin{definition}
\label{definition:coloring}
Let $c$ be a coloring function of the incidence graph $G^*$. Define\\
1.The colored edge $\vec{e}_{i,j}=(c_{i,j},c_{j,i} )$ is a {\it constant} if $c_{i,j}=c_{j,i}$; otherwise it is a {\it variable}. The number of variables is denoted by $n_c$.\\
2.{\it Vertex constraint}: The coloring function $c$ is {\it consistent} if colors assigned to those links incident to the vertex $v$ are all distinct for all $v\in V(G^*)$.	\\
3.{\it Edge constraint}: The coloring function $c$ is {\it proper} if it is consistent and all colored edges are constant.
\end{definition}

The coloring of incidence graph $G^*$ is called a {\it configuration} of graph $G$. As an example, a consistent configuration of a tetrahedron containing two variables is shown in Fig. \ref{fig:incidencetetrahedron} (b), and a proper configuration of a tetrahedron is shown in Fig. \ref{fig:incidencetetrahedron} (c) with the set of colors $C=\{g,r,b\}$, where $g$, $r$, and $b$, respectively represent green, red, and blue.

The edge constraints are sets of equalities of colors and the vertex constraints are un-equalities. The equality is a transitive binary relation but the un-equality is not. Most edge-coloring algorithms assign edge colors to satisfy vertex constraints, sets of un-equalities. However, mathematically, it is more natural and usually much easier to solve problems with equalities. Initially, we start with an arbitrary consistent color function $c$ with $\Delta$ colors that satisfies the vertex constraint at every vertex of the graph, which may contain variable-colored edges. Our edge coloring algorithm is a systematic procedure to eliminate those variables, similar to the procedure of solving a system of simultaneous equations.

\subsection{Kempe Walks}
In general, variable eliminations require a sequence of color exchanges, called {\it Kempe walks} that are performed on a two-colored {\it Kempe path} defined as follows.

\begin{definition}
\label{definition:KempePath}
In a consistently colored incidence graph, a {\it $(\alpha,\beta)$-Kempe path}, or simply {\it $(\alpha,\beta)$ path}, where $\alpha,\beta \in C$ and $\alpha \neq \beta $, is a sequence of adjacent links $l_1,l_2,...,l_{n-1},l_n$ such that $c(l_i)\in \{\alpha,\beta\}$ for $i=1,...,n$. The corresponding sequence of vertices contained in the path is defined as its {\it interior chain}. There are two types of maximum $(\alpha,\beta)$ path:\\
1. {\it $(\alpha,\beta)$ cycle} or {\it closed $(\alpha,\beta)$ path}: The two end-links $l_1$ and $l_n$ are adjacent to each other.\\
2. {\it Open $(\alpha,\beta)$ path}: Only one $\alpha$ or $\beta$ colored link is adjacent to the end-links $l_1$  and $l_n$.
\end{definition}

A $(\alpha,\beta)$-variable edge is always contained in a maximum $(\alpha,\beta)$ path, either a $(\alpha,\beta)$ cycle or an open $(\alpha,\beta)$ path. Note that the open path may either end at a fictitious vertex or a real vertex. The following lemma can be obtained by simple counting arguments.

\begin{lemma}
\label{lemma:oddeven}
An even (odd) $(\alpha,\beta)$ cycle contains even (odd) number of $(\alpha,\beta)$ variables. 
\end{lemma}
\begin{proof}
Deleting all variables  in the $(\alpha,\beta)$ cycle by edge contraction, the remaining constant edges constitute an even $(\alpha,\beta)$ cycle. The lemma is established by the following relation:
\begin{center}
 $\#$variables=$\#$edges $-$ even $\#$constant edges. \qed
\end{center}
\end{proof}

Variable eliminations can be achieved by the following color-exchange operation performed on adjacent colored edges.

\begin{definition}
\label{definition:colorexchange}
Let $\vec{e}_{j,i}  =(c_{j,i},c_{i,j} )$   and $\vec{e}_{i,k}=(c_{i,k},c_{k,i})$ be two colored edges incident to the same vertex $v_i$, written as $(c_{j,i},c_{i,j} )\circ (c_{i,k},c_{k,i} )$. Suppose that $c_{i,j}=\beta$ and $c_{i,k}=\alpha$, for $\alpha,\beta \in C=\{c_1,...,c_\Delta \}$, the binary operation $\otimes$ defined below exchanges the colors of link $l_{i,j}$ and $l_{i,k}$ incident to $v_i$:
\begin{align}
  (c_{j,i},c_{i,j} ) \otimes (c_{i,k},c_{k,i}) = (c_{j,i},\beta ) \otimes (\alpha,c_{k,i}) \Rightarrow (c_{j,i},\alpha) \circ (\beta ,c_{k,i}) \nonumber
\end{align}
\end{definition}

The binary operation $\otimes$ is non-commutative but associative, it can be considered as a transformation of a consistent coloring function $c$ to another consistent coloring function $c^{'}$ such that $c_{i,j}^{'}=c_{i,k}=\alpha$ and $c_{i,k}^{'}=c_{i,j}=\beta$. Two adjacent variables may be eliminated by the color-exchange operation. For example, if the two colored edges are variables $\vec{e}_{j,i}=(\alpha,\beta)$ and $\vec{e}_{i,k}=(\alpha,\gamma)$, then the color exchange $\vec{e}_{j,i} \otimes  \vec{e}_{i,k} = (\alpha,\beta) \otimes (\alpha,\gamma) \Rightarrow (\alpha,\alpha) \circ (\beta,\gamma) $ can eliminate one of these variables. On the other hand, color exchanges between two adjacent constant edges may introduce new variables. A color-exchange operation is {\it effective} if it does not increase the number of variables $n_c$.

\begin{figure}[t]
	\centering
	\includegraphics[scale=0.85]{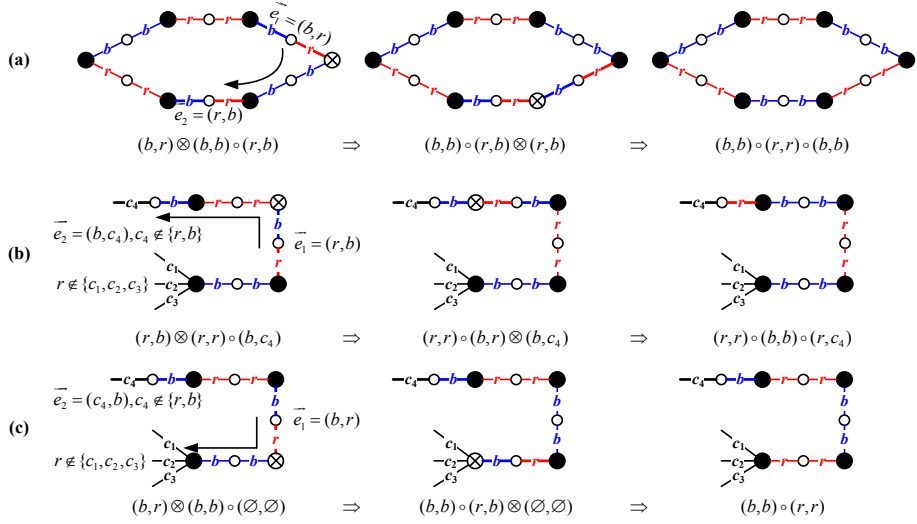}
	\caption{Variable eliminations by Kempe walks.}
	\label{fig:KempeWalk}
\end{figure}

The Kempe walk of a $(\alpha,\beta)$ variable on a $(\alpha,\beta)$ path is a sequence of effective color-exchange operations performed on its interior chain. Examples of variable eliminations by Kempe walks are provided in Fig. \ref{fig:KempeWalk}. Consider the $(r,b)$ path $(b,r)\circ (b,b) \circ(r,b)$ shown in Fig. \ref{fig:KempeWalk} (a), the variable $\vec{e}_1=(b,r)$ can walk to another variable $\vec{e}_2=(r,b)$ by the following sequence of color exchanges performed on its interior chain:
\begin{align}
 (b,r) \otimes  (b,b) \circ (r,b) \Rightarrow (b,b)\circ (r,b) \otimes (r,b) \Rightarrow (b,b)\circ (r,r) \circ (b,b), \nonumber
\end{align}
in which two variables are eliminated by color exchanges.

In a regular graph $G$, walks on an open $(\alpha,\beta)$ path always terminate on fictitious vertices at both ends; however, they may terminate on a vertex $v_i\in V(G)$ with a missing $\alpha$ or $\beta$ link if the graph $G$ is irregular. A missing colored edge at a vertex can be regarded as a {\it don't care edge}, denoted as $(\emptyset,\emptyset)$. At those degenerate vertices, the color-exchange operation involving a don't care edge is symbolically expressed as $(\alpha,\beta) \otimes (\emptyset,\emptyset) \Rightarrow (\alpha,\alpha)$. The examples depicted in Fig. \ref{fig:KempeWalk} (b) and (c) show that a $(b,r)$ variable on an open $(b,r)$ path can be eliminated by walking to either end of the path, regardless if it is a vertex or a fictitious vertex.

Table \ref{tab:KempeWalkcase} summarizes all possible one-step Kempe walks of a $(\alpha,\beta)$ variable. Note that cases KW3 and KW4 only occur at the end of an open $(\alpha,\beta)$ path, and the number of variables $n_c$ monotonically decreases as long as the $(\alpha,\beta)$ variable is walking on a $(\alpha,\beta)$ path.

\begin{table}[htbp]
	\centering
	{\footnotesize
		\begin{tabular}{|c |c| c| c| }
			\hline 
			Case & Next Step & Operation & Result \\
			\hline \hline
			KW1 & $(\alpha,\beta)\circ (\alpha,\alpha)$ & $(\alpha,\beta)\otimes(\alpha,\alpha)\Rightarrow(\alpha,\alpha)\circ(\beta,\alpha)$&step forward\\ \hline 
			KW2 & $(\alpha,\beta)\circ (\alpha,\beta)$ & $(\alpha,\beta)\otimes(\alpha,\beta)\Rightarrow(\alpha,\alpha)\circ(\beta,\beta)$& eliminate two variables \\ \hline 
			KW3 & $(\alpha,\beta)\circ (\alpha,\gamma)$ & $(\alpha,\beta)\otimes(\alpha,\gamma)\Rightarrow(\alpha,\alpha)\circ(\beta,\gamma)$& eliminate one variable \\ \hline 
			KW4 & $(\alpha,\beta)\circ (\emptyset,\emptyset)$ & $(\alpha,\beta)\otimes(\emptyset,\emptyset)\Rightarrow(\alpha,\alpha)$& eliminate one variable \\
			\hline		
		\end{tabular}	
		}
	\caption{One-step move of $(\alpha,\beta)$ variable on $(\alpha,\beta)$ path.}
	\label{tab:KempeWalkcase}
\end{table}

\subsection{Color Quaternion}
Define the {\it negation} of a colored variable $(\alpha,\beta)$, denoted by $-(\alpha,\beta)=(\beta,\alpha)$, as the color vector in the opposite direction of $(\alpha,\beta)$. The following color-inversion operation can change the direction of a $(\alpha,\beta)$ variable contained in a maximum $(\alpha,\beta)$ path.

\begin{definition}
\label{definition:colorinversion}
The {\it color inversion} performed on a maximum $(\alpha,\beta)$ path $H$ exchanges color $\alpha$ and $\beta$ on all links of $H$.
\end{definition}

The color inversion operation requires a sequence of color exchanges involving all vertices of the interior chain of the $(\alpha,\beta)$ path. It can only apply to a maximum $(\alpha,\beta)$ path, either a $(\alpha,\beta)$ cycle or an open $(\alpha,\beta)$ path, to avoid increasing the number of variables $n_c$. For consistency, we also use the notation $-(\alpha,\alpha)=(\alpha,\alpha)(\bmod 2)$ to denote the negation of a constant edge. With the notion of negation, the color-exchange operation of complex colors follows the same multiplication rules as quaternion.

The concept of quaternion was introduced by Hamilton in 1843 \cite{conway2003quaternions} as an extension of complex numbers. The basis elements of quaternions are customarily denoted as 1, $i$, $j$, and $k$. The correspondences between quaternion multiplication and color-exchange operations are summarized in Table \ref{tab:Quaternion}. Note that the minus sign of diagonal entries in the color-exchange table indicates the necessary color inversion of the right operand due to the consistency requirement at each vertex of the color configuration. The exchange operation listed in Table \ref{tab:Quaternion} implies that if two adjacent variables have one color in common, then at least one of them can be eliminated. This is an important property in the construction of edge-coloring algorithms for eliminating variables involving more than three colors.

\begin{table}[htbp]
	\centering
{\footnotesize
\begin{tabular}{|c|c|} \hline
\multicolumn{2}{|c|}{Correspondence} \\ \hline
\multicolumn{2}{|c|}{$A=(\alpha,\alpha)=-(\alpha,\alpha),B=(\beta,\beta)=-(\beta,\beta),C=(\gamma,\gamma)=-(\gamma,\gamma) \rightarrow 1 $
} \\
\multicolumn{2}{|c|}{$(\alpha,\beta) \rightarrow i,( \beta,\alpha) \rightarrow -i, (\beta,\gamma) \rightarrow j,(\gamma,\beta ) \rightarrow -j,	(\gamma,\alpha ) \rightarrow k, (\alpha, \gamma) \rightarrow -k$
} 
\\ \hline

Quaternion Multiplication & Color-exchange Operation \\ \hline
$\left[ \begin{array}{cccc} \times & -i & -j & -k \\ 
			                             i & 1&-k&j \\ 
													    			j& k &1 &-i\\
                              			k& -j& i & 1\\
																		\end{array} \right]$ & 
																		
																		$\left[ \begin{array}{cccc} \otimes & (\beta,\alpha) & (\gamma,\beta) & (\alpha,\gamma) \\ 
			                             (\alpha,\beta) & -AB&(\alpha,\gamma)B&A(\beta,\gamma) \\ 
														       (\beta,\gamma) & B(\gamma,\alpha)&-BC&(\beta,\alpha)C \\
														      (\gamma,\alpha) & (\gamma,\beta)A&C(\alpha,\beta)&-CA \\
																	\end{array} \right]$ \\ 
\hline
\end{tabular}
}
\caption{Correspondence between quaternion multiplication and color-exchange operations.}
	\label{tab:Quaternion}
\end{table}

\section{Canonical Configurations}
Kempe walks provide the most efficient and natural way to eliminate variables. Almost all variables in an arbitrary initial color configuration can be eliminated by implementing Kempe walks. A configuration is called a {\it canonical configuration} if no variables can be further reduced by single variable Kempe walks. In a canonical configuration, it is easy to show from Lemma \ref{lemma:oddeven} that all remaining variables are contained in odd cycles, and every odd cycle only contains a single variable. The following algorithm exhaustively eliminates variables by Kempe walks.

\begin{table}[htbp]
	\centering
	{\footnotesize
			\begin{tabular}{|l|l| } \hline
			\multicolumn{2}{|c|}{Walk-on-Kempe-Path Algorithm (WKP)} \\ \hline
			{\bf Algorithm:} Walk-on-Kempe-Path (WKP)                        &  {\bf Subroutine:} Variable-Walk \\
 			{\bf Input:} an initial $\Delta$-edge-coloring configuration of $G^*$ &  {\bf Input:} variable $\vec{e}_{i,j}=(\alpha,\beta)$ \\
 			{\bf Output:} a proper $\Delta$-edge-coloring or a canonical            &  {\bf Output:} return true if $\vec{e}_{i,j}$ is eliminated;\\
			 configuration of $G^*$                           &  otherwise, return false \\            
			1. {\it VariableList} $\leftarrow$ find all variables            &  9. ~{\it KempePath} $\leftarrow$ staring from one end  of \\
			2. {\bf if}  {\it VariableList} is empty {\bf then}              &  \quad ~ $\vec{e}_{i,j}$, say $v_i$, search for $(\alpha,\beta)$ path until  \\
			\quad {\bf return} a properly colored graph                      &  \quad ~(1) it finds another variable   \\
			3. {\bf else}                                                    &  \quad ~(2) it terminates at a vertex with missing\\ 
			4. \quad {\bf for each} variable $\vec{e}$ in {\it VariableList} {\bf do}  & \quad ~ $\alpha$ or $\beta$ link\\
			5. \quad \quad {\bf if} Variable-Walk($\vec{e}$)=true            &  \quad ~(3) it returns to $v_j$, thus forms a cycle \\
			6. \quad \quad {\bf then} update {\it VariableList} and {\bf goto} step 2 &  10.	{\bf if} case (3) occurred, {\bf then return} false \\
			7. \quad \quad {\bf else} continue 															 &  11. {\bf else} color exchange on interior chain of \\
			8. {\bf return} a canonical configuration                        &	\quad ~{\it KempePath} and {\bf return} true\\	
			\hline		
		\end{tabular}	
		}
	\label{algo:WKPAlgo}
\end{table}

\begin{theorem}
\label{theorem:canonical}
For graph $G=(V,E)$, the algorithm WKP either returns a proper $\Delta$-edge-coloring or a canonical configuration. The running time of the algorithm is on the order of $O(|V||E|^2)$
\end{theorem}

\begin{proof}
Because the subroutine {\it Variable-Walk} returns false at step 10 only when case (3) occurs, according to Lemma \ref{lemma:oddeven}, the input variable $\vec{ e}_{i,j}=(\alpha,\beta)$ must be the only variable contained in an odd $(\alpha,\beta)$ cycle. The algorithm returns a properly colored graph in step 2 if the {\it VariableList} is empty. Otherwise, every variable in the {\it VariableList} is contained in an odd Kempe cycle, and the algorithm returns a canonical configuration at step 8.

Next, the order of running time $O(|V| |E|^2)$ can be estimated from the number of times that the subroutine {\it Variable-Walk} is executed, and the running time of the subroutine {\it Variable-Walk}. The number of times that the subroutine {\it Variable-Walk} is executed is a function of $n_c$, the initial number of variables, denoted as $\Phi(n_c)$. Since at least one variable is eliminated in an updated {\it VariableList} when the subroutine repeats in the loop (steps 4-7), then $\Phi(n_c) \leq \sum_{i=0}^{n_c}(n_c -i) \in O({n_c}^{2})\subset O(|E|^2) $. The running time of the subroutine {\it Variable-Walk} is on the order of $O(|V|)$, because both the running time of path searching in step 9 and the number color-exchange operations performed in step 11 are bounded by the number of vertices $|V|$. \qed
\end{proof}

An immediate consequence of the above theorem is the following corollary.
\begin{corollary}
For bipartite graph $G=(V,E)$, the WKP algorithm always returns a proper $\Delta$-edge-coloring in $O(|E|\log{|V|})$ time.
\end{corollary}

\begin{proof}
Since there are no odd cycles in bipartite graphs, then the subroutine {\it Variable-Walk} always eliminates at least one variable and the subroutine is executed at most $n_c$ times, where $n_c$ is the initial number of variables. Let $L_i$  be the length of the search path of the $i$th input variable in step 9. Since the number of color exchanges and the path search time are proportional to the total path length $\sum_{i=1}^{n_c}L_i$ , the average path length in a random graph is on the order of $O(\log{|V|})$ \cite{fronczak2004average}. Thus, the running time of the {\it WKP} algorithm for a bipartite graph $G=(V,E)$  is bounded by 
\begin{align}
\sum_{i=1}^{n_c}L_i =n_c \frac{\sum_{i=1}^{n_c}L_i}{n_c} \in  n_c O(\log{|V|} )\subset O(|E|\log{|V|} ) \nonumber \qed
\end{align} 
\end{proof}
The above bound is a conservative estimate, because we did not take the trade-off between the number of variables $n_c $ and the path lengths in a bipartite graph into consideration. 

\begin{figure}[htpb]
	\centering
	\includegraphics[scale=0.80]{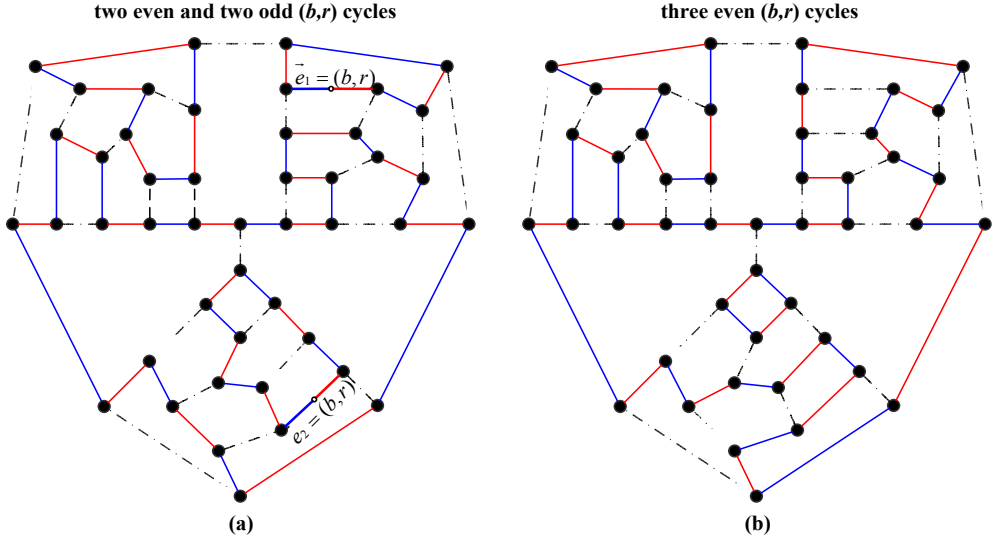}
	\caption{3-edge-colored Tutte graph.}
	\label{fig:Tutte}
\end{figure}

For non-bipartite graphs, two examples of the 3-edge-colored Tutte graphs that resulted from the {\it WKP} algorithm are depicted in Fig. \ref{fig:Tutte}, where green-colored edges and fictitious vertices are faded to highlight $(b,r)$ cycles. The canonical configuration shown in Fig. \ref{fig:Tutte} (a) has two remaining $(b,r)$ variables, respectively, contained in two disjoint odd $(b,r)$ cycles. Fig. \ref{fig:Tutte} (b) shows that a properly colored Tutte graph has three even $(b,r)$ cycles.

\section{Walks on Directional Paths}

Despite the fact that an overwhelming number of variables in the initial color configuration can be eliminated by Kempe walks, the graph $G$ is still not properly colored if some variables remain trapped in odd Kempe cycles. The Kempe walks are limited to color exchanges performed within alternating paths, which are fixed subgraphs $H\subset G$ in any given color configuration. In this section, we introduce walks on directional paths, which involve more than two colors, to systematically eliminate remaining variables in a canonical configuration.  

Consider that a tagged variable $\vec{e}_1=(\alpha,\beta)$  walks along a predetermined directional path. A move of  $\vec{e}_1$  is effective if either the tagged variable can step forward, or some variables are eliminated along the way. If $\vec{e}_2=(\gamma,\delta)$, where $\gamma \neq \alpha $ and $\delta \neq \beta$, is the next edge adjacent to $\vec{e}_1$ on the path, then it is necessary to change the colors of $\vec{e}_2$, either $\gamma$ or $\delta$, such that the operation $\vec{e}_1 \otimes \vec{e}_2$ yields an effective move of $\vec{e}_1$. That is, effective walks on directional paths are actually walks on dynamically changed Kempe paths. As indicated in Table \ref{tab:Quaternion} in color quaternion multiplication, any effective moves only involve three colors. Thus, there are only two useful types of color inversion on the next edge $\vec{e}_2$: 
\begin{align}
\alpha \mbox{-type}: \vec{e}_2 & =(\gamma,\delta) \rightarrow \vec{e}_2=(\alpha,*) \nonumber \\ 
\beta \mbox{-type}: \vec{e}_2 & =(\gamma,\delta) \rightarrow \vec{e}_2=(*,\beta) \nonumber 
\end{align}
Note that a color inversion operation applied to $\vec{e}_2$ may become invalid if the operation also involve $\vec{e}_1$. If a valid color conversion cannot be found, then the tagged variable $\vec{e}_1$ is {\it blocked}. All possible one-step non-Kempe moves of a tagged variable on a directional path in a canonical configuration are given in the Appendix.

It should be expected that blocking may occur along the way; otherwise, variables can all be eliminated by walking them to a common vertex, resulting in a proper $\Delta$-edge-colored graph. The Petersen graph is a well-known counterexample to show that this is impossible. In the reminder of this section, we describe a variable elimination algorithm by walks on a directional spanning tree with deflections.

\subsection{Walks on Spanning Trees with Deflections}
Intuitively, a spanning tree of the graph $G$ can provide efficient directional paths that guide remaining variables in canonical configurations to walk to a common destination. In view of the analogy between variable edges and vectors, we can think of the paths of a spanning tree as {\it coordinates} of a vector space embedded in the graph $G$ with the origin at the root of the tree. Presumably, if variables can freely walk on those directional paths, then they will either intercept each other on the way or eventually meet at the root. A blocked variable can be randomly deflected to another nearby directional path and resume the walking toward the root. The detailed steps of variable eliminations are listed in the {\it WST} algorithm.
\begin{table}[htbp]
	\centering
	{\footnotesize
			\begin{tabular}{|l|l| }		 \hline
			\multicolumn{2}{|c|}{Walk-on-Spanning-Tree Algorithm (WST) } \\ \hline
		  {\bf Algorithm:} Walk-on-Spanning-Tree(WST)    						 			 & {\bf Subroutine:} Walk-to-Next-Step 	\\	
			{\bf Input:} incidence graph $G^*$										  						 			 & {\bf Input:} variable $\vec{e}_{i,j}=(\alpha,\beta)$, spanning tree {\it ST},	\\	
			{\bf Output:} proper $\Delta$-edge-coloring or claim $\chi_e(G)=\Delta+1$ & color $c_i\in\{\alpha,\beta$\}	\\	
			1.~ {\it ST} $\leftarrow$ construct a spanning tree of $G$	           		 & {\bf Output:} return true if variable containing $c_i$   	\\	
			2.~ initial coloring of $G^*$																							 & is eliminated; otherwise, variable moves to  \\
			3.~ execute {\it KWP} algorithm on $G^*$                                         & the next edge and returns false\\ 
			~ \quad ({\bf comment:} results in proper $\Delta$-edge-coloring,  or a    & 16. pick a next edge $\vec{e}_{l,k}$ toward the root of {\it ST}	\\
			~ \quad canonical configuration according to theorem \ref{theorem:canonical}) & 17. {\bf loop until} $\vec{e}_{i,j} \rightarrow \vec{e}_{l,k}$ succeeds	\\
			4.~ {\bf for each} color $c_i$ in $C=\{c_1,...,c_{\Delta}\}$              &18. \quad $\vec{e}_{l,k} \leftarrow $ pick a fresh adjacent edge	\\
			5.~ \quad {\it VariableList} $\leftarrow$ all variable edges containing $c_i$  & 19. {\bf end loop} \\
		  6.~ \quad {\bf if} {\it VariableList} is empty, {\bf then goto} step 4 and  & 20. {\bf if} $\vec{e}_{i,j} \rightarrow \vec{e}_{l,k}$ eliminates variable   \\ 
			~\quad \quad  continue with next color $c_{i+1}$                           & ~ \quad containing $c_i$ or Variable-Walk($\vec{e}_{l,k}$)=true  \\
			7.~ \quad {\bf else}                                                       & ~ \quad {\bf then return} true\\
			8.~ \quad \quad {\bf for each} variable $\vec{e}$ in {\it VariableList}    & ~ \quad 	({\bf comment:} execute  Variable-Walk  \\
			9.~ \quad \quad \quad {\bf loop} $r(n_i,k)-1$ times                          & ~ \quad  subroutine in WKP algorithm in case \\
			10. \quad \quad \quad \quad {\bf if} Walk-to-Next-Step($\vec{e},ST,c_i$)=true & ~ \quad  new Kempe paths are created)\\
			\quad \quad \quad \quad \quad ~~{\bf then} update {\it VariableList}, go to step 6 & 21. {\bf else return} false \\
			11. \quad \quad \quad  {\bf end loop}                                      & \\
			12. \quad \quad  {\bf end for each}                                        &\\
			13. \quad \quad  {\bf output} $\chi_e(G)=\Delta+1$                         &\\
			14. {\bf end for each}                                                     &\\
			15. {\bf output} proper $\Delta$-edge-coloring of $G$                      &\\
			
			\hline		
		\end{tabular}	
		}
	\label{tab:voice-languages}
\end{table}

The  {\it WST} algorithm is initialized by a spanning tree and an arbitrary color configuration of $G$ with a set of ${\Delta}$ colors $C=\{c_1,...,c_\Delta \}$. In the canonical configuration generated in step 3, we firstly eliminate variables that contain color $c_1$, then eliminate the remaining variables that contain color $c_2$, and so on. The process is similar to the Gauss elimination for solving systems of linear equations, in which variables are eliminated one kind at a time. 

In the $i$th iteration started at step 4, the {\it VariableList} only includes those variables containing color $c_i$, sometime called $c_i$-{\it variables}. In steps 8-12, each selected $c_i$-variable is allowed to walk towards the root for $r(n_i,k)-1$ steps, where the parameter $r(n_i,k)$ is a function to be determined later in the next subsection. If the subroutine {\it Walk-to-Next-Step} returns true, indicating the elimination of a $c_i$-variable, then another $c_i$-variable will be selected from the {\it VariableList} and the above process is repeated. If the {\it VariableList} is empty, the algorithm goes back to step 4 and focuses on the next color $c_{i+1}$. The problem is solved if all variables are eliminated and a proper $\Delta$-edge-coloring is reached. On the other hand, if a $c_i$-variable in the $i$th iteration fails to eliminate any $c_i$-variables within  $r(n_i,k)-1$ steps, then the algorithm halts in step 13 and claims that $\chi_e(G)=\Delta+1$. 

In step 16 of the subroutine {\it Walk-to-Next-Step}, we first choose an adjacent edge $\vec{e}_{l,k}$ that is closer to the root. If the move to next edge $\vec{e}_{i,j} \rightarrow \vec{e}_{l,k}$ is blocked, then we choose another adjacent edge until it succeeds in steps 17-19. Since a blocked variable $\vec{e}_{i,j}=(\alpha,\beta)$ can always be deflected to one of the two neighboring $\alpha$ or $\beta$ links, there is no risk of running into an infinite loop in steps 17-19 even in the worst-case scenario. 

\subsection{Analysis of Randomized Algorithm}

In general, walks on carefully selected paths, such as spanning tree, are more efficient than random walks. However, the efficiency of guided walks depends on path selection, which makes the complexity analysis mathematically intractable. On the other hand, the complexity of path-independent random walks is easier to estimate, and it provides an upper bound of the complexity of all guided variable elimination methods. 

The basic idea of the random-walk algorithm is similar to that of the {\it WST} algorithm, except in step 16, the neighboring edge $\vec{e}_{l,k}$  of the next edge is randomly chosen. In the following analysis, we assume that, for randomly generated graph, the probability of the next edge, resulting from the execution of the subroutine {\it Walk-to-Next-Step}, is evenly distributed among all neighboring edges. If the chromatic index of the input graph is $\chi_e(G)=\Delta $ and the parameter $r(n_i,k)$ in step 9 is properly chosen, we prove that the random-walk algorithm returns a proper $\Delta$-edge-coloring in polynomial time with a probability of at least 1/2. 

Consider a random walk that starts at a vertex $v \in V$ of an $n$-vertices graph $G=(V,E)$, and whenever it reaches any vertex $u \in V$, chooses an edge at random from those edges incident to $u$, and traverses it. Suppose the random walk starts from a vertex  $v \in V$, the {\it access time} or {\it hitting time} $H[v,u]$ is the expected number of steps before vertex $u$ is visited. An $O(n^3)$ upper bound on the access time was first obtained by Aleliunas, Karp, Lipton, Lov{\'a}sz, and Rackoff \cite{Aleliunas1979random}. Later, it was proved in \cite{lovsz1993random} that the access time is at most $2n^2$ for a regular graph. Let $A\subset V$ be a proper subset of vertices and let $v\in A^c$, where $A^c=V \setminus A$, the access time $H[v,A]$ is the expected number of steps before any vertex $u\in A$ is visited, starting from vertex $v$. A proof of the following bound is given in \cite{aldous2002reversible}.

\begin{lemma} 
\label{lemma:hittime}
Consider random walk on a regular graph $G=(V,E)$. Let $A\in V$ and $v\in A^c$, then       
\begin{align}
 H[v,A]<4|A^c |^2 \nonumber 
\end{align}
\end{lemma}

To be consistent with the above analysis of random walks on graphs, we consider that the random-walk algorithm is implemented on the line graph $ \widehat{G} = (V^*,\widehat{E})$ induced from incidence graph $G^*=(V,E)$. The set of vertices $V^*$  is the set of fictitious vertices of $G^*=(V,E)$, and an ``edge'' connecting two ``vertices'' $e_i^*, e_j^*\in V^*$ if and only if $e_i$ and $e_j$  are incident to the same vertex in $G$. Moreover, we assume, without loss of generality, that the input graph is $\Delta$-regular. The running time of the random walk searching algorithm is determined by the total number of moves required for finding a proper configuration. A move of a variable in the algorithm is a transformation from one configuration to another configuration. Since any simple graph $H$ with maximum degree $\Delta$ is a subgraph of a $\Delta$-regular graph $G$ with the same number of vertices, and those edges in $G$ missing in $H$ are all don't care edges. The $\Delta$-regular graph $G$ admits far more color configurations than its subgraph $H$. Thus, we are considering the worst case in our analysis. This point is verified by the experimental results described in section 5. 

Let $n_i$ be the number of $c_i$-variables at the beginning of the $i$th iteration of the random-walk algorithm, corresponding to step 4 in the {\it WST} algorithm, and let $h(n_i,k)$ be a function defined by:
\begin{align}
\label{eq:h}
  h(n_i,k)= \frac{4 \Delta n_i(|E|-k+1)^2}{\ln{2} },
\end{align}
where $k=1,...,n_i$  is a parameter representing the number of $c_i$-variables in the {\it VariableList}. We show that the random-walk algorithm has the following property.

\begin{theorem} 
If input to the random-walk algorithm is a $\Delta$-edge-colorable graph $G=(V,E)$ and the parameter in step 9 of the algorithm is $r(n_i,k)=\left\lceil h(n_i,k) \right\rceil $, then the algorithm returns a proper $\Delta$-edge-coloring of $G$ with a probability of at least 1/2. 
\end{theorem}
\begin{proof}
We first prove that 
\begin{align}
   \mbox{Pr} \{ \mbox{all } c_i \mbox{-variables are eliminated} \} > {\left(\frac{1}{2}\right)}^{1/\Delta},  
	\end{align}
in the $i$th iteration of the algorithm. Let $A_k=\{\epsilon_1,...,\epsilon_k\}$ be the set of $k$ $c_i$-variables in the {\it VariableList}. Suppose that the variable $\epsilon_j$ takes $T_j^{(k)}$ steps to hit another variable in the set $A_k \setminus \{\epsilon_j\}$. It follows from Lemma \ref{lemma:hittime} that the access time is bounded by
\begin{align}
\label{eq:ET}
  E\left[ T_{j}^{(k)}  \right]= H \left[ \epsilon_i, A_k\setminus \{\epsilon_i\}  \right] < 4(|E|-k+1)^2,
\end{align}
for $j=1,...,k$. Since $T_j^{(k) }$ are i.i.d. random variables, the probability that no variables in the set $A_k$ can be eliminated is given by
\begin{align}
\label{eq:Akfails}
   \mbox{Pr} \{ A_k \mbox{ fails to reduce to } A_{k-1} \} &=\mbox{Pr} \left\{ T_{1}^{(k)} \geq r(n_i,k),..., T_{k}^{(k)}\geq r(n_i,k) \right\} \nonumber  \\
	&= \mbox{Pr}\left\{ T_{j}^{(k)} \geq r(n_i,k)\right\}^k
\end{align} 
From Markov inequality we have
\begin{align}
\label{eq:makoveieq}
\mbox{Pr}\left\{ T_{j}^{(k)} \geq r(n_i,k) \right\}^k &\leq \left( \frac{E\left[ T_{j}^{(k)} \right]}{r(n_i,k)}\right) ^k  < \left(\frac{ H \left[ \epsilon_i, A_k\setminus \{\epsilon_i\}  \right]}{h(n_i,k)}  \right) ^k < \left(  \frac{\ln{2}}{\Delta n_i} \right)^k.
\end{align}
The probability that all $c_i$-variables can be eliminated in the $i$th iteration of the algorithm is given by
\begin{align}
\label{eq:allelimiated}
\mbox{Pr}\left\{ \mbox{all } c_i \mbox{-variables are eliminated} \right\} =  \prod_{k=1}^{n_i} (1- \mbox{Pr} \{ A_k \mbox{ fails to reduce to } A_{k-1} \}).
\end{align}
Substituting (\ref{eq:Akfails}) and (\ref{eq:makoveieq}) into (\ref{eq:allelimiated}), we have
\begin{align}
\label{eq:allelimiatednumber}
\mbox{Pr}\left\{ \mbox{all } c_i \mbox{-variables are eliminated} \right\} &> \prod_{k=1}^{n_i} \left( 1- \left(  \frac{\ln{2}}{\Delta n_i} \right)^k \right)  > \prod_{k=1}^{n_i} \left( 1- \left(  \frac{\ln{2}}{\Delta n_i} \right) \right)  \nonumber \\
&= \left( 1- \left(  \frac{\ln{2}}{\Delta n_i} \right) \right) ^{n_i} \sim e^{-\ln{2}/\Delta} = \left(\frac{1}{2} \right)  ^{1/\Delta}.
\end{align}
If $\chi_e(G)=\Delta$, then the probability that the algorithm returns a proper $\Delta$-edge-coloring of $G$ is given by
\begin{align}
\mbox{Pr} \left\{ \mbox{output a proper $\Delta$-edge-coloring}  | \chi_e(G)=\Delta \right\} &= \prod_{i=1}^{\Delta} \mbox{Pr}\left\{ \mbox{all } c_i \mbox{-variables are eliminated} \right\}  \nonumber \\
&>  \prod_{i=1}^{\Delta}  \left(\frac{1}{2} \right)  ^{1/\Delta} =1/2 \nonumber \qed
\end{align}
\end{proof}

\begin{theorem} 
The running time of the random-walk algorithm with $r(n_i,k)=\left\lceil h(n_i,k) \right\rceil$ is $O(\Delta|V||E|^5)$.
\end{theorem}
\begin{proof}
The number of steps $\Phi_i$ for eliminating all $c_i$-variables is bounded by
\end{proof}
\begin{align}
 \Phi (c_i) &\leq \sum_{k=1}^{n_i}k(r(n_i,k)-1) < \sum_{k=1}^{n_i}kh(n_i,k) =  \frac{4\Delta n_i}{\ln{2}}\sum_{k=1}^{n_i}k(|E|-k+1)^2  \nonumber \\
&\leq  \frac{4\Delta n_i}{\ln{2}} \sum_{k=1}^{n_i}k|E|^2 = \frac{2\Delta n_i|E|^2n_i(n_i+1)}{\ln{2}}.
\end{align}
Therefore, the total number of steps $\Phi$ after eliminating all variables is bounded by
\begin{align}
\Phi = \sum_{i=1}^{\Delta} \Phi (c_i) < \frac{2\Delta |E|^2}{\ln{2}}  \sum_{i=1}^{\Delta}({n_i}^{3}+{n_i}^{2}) .
\end{align} 
The total number of variables is monotonic decreasing throughout the entire process, which implies $\sum_{i=1}^{\Delta}n_i \leq n_c$. Then we have
\begin{align}
\Phi <\frac{2\Delta |E|^2}{\ln{2}}   \sum_{i=1}^{\Delta}({n_i}^{3}+{n_i}^{2}) \leq \frac{2\Delta |E|^2}{\ln{2}}({n_c}^{3}+{n_c}^{2}) . 
\end{align} 
Hence, the complexity in terms of the number of steps is bounded by 
\begin{align}
\Phi \in O(\Delta |E|^2{n_c}^3)\subset O(\Delta |E|^5). 
\end{align}
From the proof of theorem \ref{theorem:canonical}, we know that the running time of the subroutine {\it Variable-Walk}, or the running time of each successful move, is on the order of $O(|V|)$. Therefore, the running time of random walk coloring algorithm is on the order of $O(\Delta|V||E|^5)$.\qed

\subsection{Snarks}
\begin{figure}[htpb]
	\centering
	\includegraphics[scale=0.8]{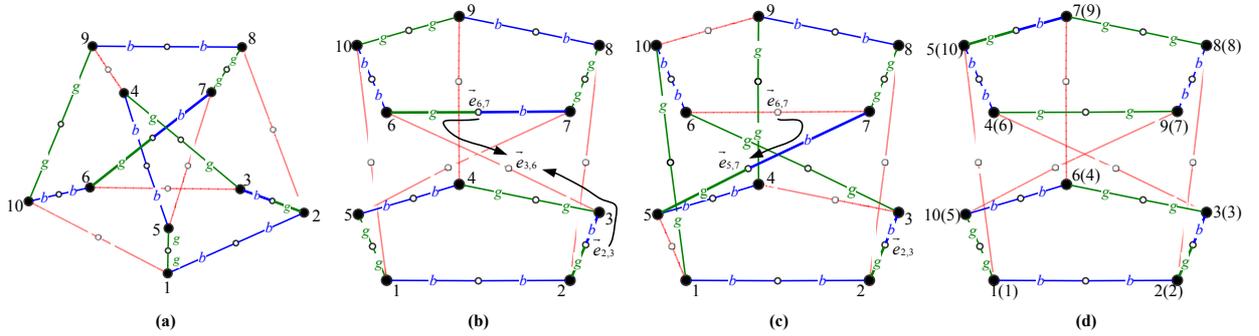}
	\caption{Isomorphic configurations of Petersen graph (red color is faded). (a) Canonical configuration. (b) Twin-cycle configuration, $\vec{e}_{6,7} \rightarrow \vec{e}_{3,6}$ and $\vec{e}_{2,3} \rightarrow \vec{e}_{3,6}$ are blocked. (c) $\vec{e}_{6,7} \rightarrow \vec{e}_{5,7}$ succeed. (d) Twin-cycle configuration after $\vec{e}_{6,7} \rightarrow \vec{e}_{5,7}$.}
	\label{fig:snark}
\end{figure}

The main difference between solving linear equations and edge coloring is the recognizing of final state. The inconsistency of a system of linear equations can be easily identified by variable eliminations in polynomial time. But eliminating variables in a class 2 graph may run into an infinite loop. The smallest class 2 cubic graph, called {\it snark}, is Petersen graph. The 3-color canonical configuration of the Petersen graph shown in Fig. \ref{fig:snark} (a) contains two variables in two disjoint odd cycles. The two odd cycles behave the same as two parallel lines in a Euclidean space; they can never cross each other, which is the geometric interpretation of inconsistent linear equations.
 
In general, all 3-color canonical configurations of Petersen graph are isomorphic. That is, corresponding to any maximum $(\alpha,\beta)$ path $H_1$  in a canonical configuration $\phi_1$, there is a maximum $(\alpha,\beta)$ path $H_2$  in another canonical configuration $\phi_2$, such that the two subgraphs $H_1$ and $H_2$ are graph isomorphic \cite{west2001introduction}. This is the reason that any variable elimination procedures can never halt when all canonical configurations of graph $G$ are isomorphic. 

The configuration shown in Fig. \ref{fig:snark} (b) is the same as that in Fig. \ref{fig:snark} (a), but the two disjoint $(b,g)$ cycles are separated in the plane, and connected by constant $(r,r)$ edges. In Fig. \ref{fig:snark} (b), we consider the walk on the shortest directional path $\vec{e}_{6,7} \circ \vec{e}_{3,6} \circ \vec{e}_{2,3}=(b,g)\circ (r,r)\circ (b,g)$ between the two variables $\vec{e}_{6,7}$  and  $\vec{e}_{2,3}$. Both moves, $\vec{e}_{6,7} \rightarrow \vec{e}_{3,6}$ and $\vec{e}_{2,3} \rightarrow \vec{e}_{3,6}$ are blocked according to the blocked move of DW1.2 described in Table \ref{tab:nonKempeWalkcase}(see the Appendix). In fact, walks on any shortest directional path will be blocked if the two variables are directly connected by a constant $(r,r)$ edge. 

Fig. \ref{fig:snark} (c) shows that the variable $\vec{e}_{6,7}$ successfully walks to edge $\vec{e}_{5,7}$. However, the resulting configuration is the same as the previous one up to some permutation of vertices. The correspondence between the vertices in Fig. \ref{fig:snark} (b) and Fig. \ref{fig:snark} (c) is shown in Fig. \ref{fig:snark} (d). Therefore, whenever a blocked variable walks out of the odd cycle, the new configuration is isomorphic to the previous one.

The edge coloring of general graphs also faces the halting problem. In fact, any $\Delta$-regular graph $G$ with an odd number of vertices is a class 2 graph. A simple example is $K_5$, where $n=5, \Delta=4$ and $\chi_e(K_5 )=5$. In a 4-color canonical configuration of $K_5$, two distinct variables are mutually blocked and can never be eliminated, which corresponds to the blocked move of DW2.2 given in Table \ref{tab:nonKempeWalkcase} (see the Appendix).

\section{Experimental Study}

Experiments were conducted on randomly generated graphs and selected benchmark graphs. The results show that the performance of our algorithm on random graphs agrees with theoretical expectations, and our algorithm can efficiently solve many hard instances. All experiments ran on a Linux KVM virtual machine with one CPU core at 2.66 GHz and 512 MB memory. The coloring algorithm was implemented in C and compiled by GCC 4.4.3. Python scripts were used to generate random graph instances as well as manage the experiments.
	
\subsection{Results of Random Graphs}
\begin{figure}[htbp]
		\centering
		\subfloat[4-regular $(\Delta=4, \delta=2)$]{\includegraphics[width=0.45\textwidth]{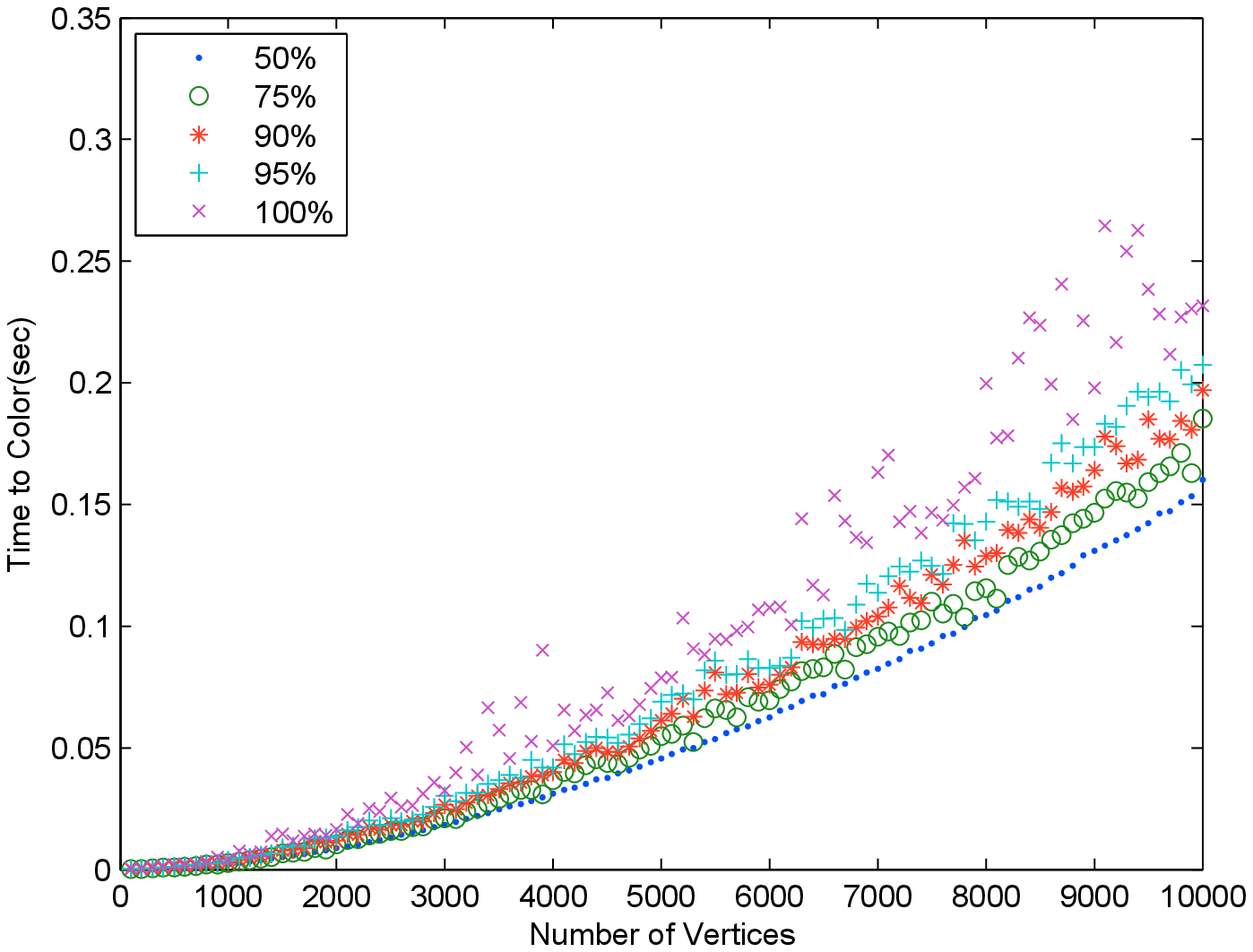}}
		\subfloat[8-regular $(\Delta=8, \delta=4)$]{\includegraphics[width=0.45\textwidth]{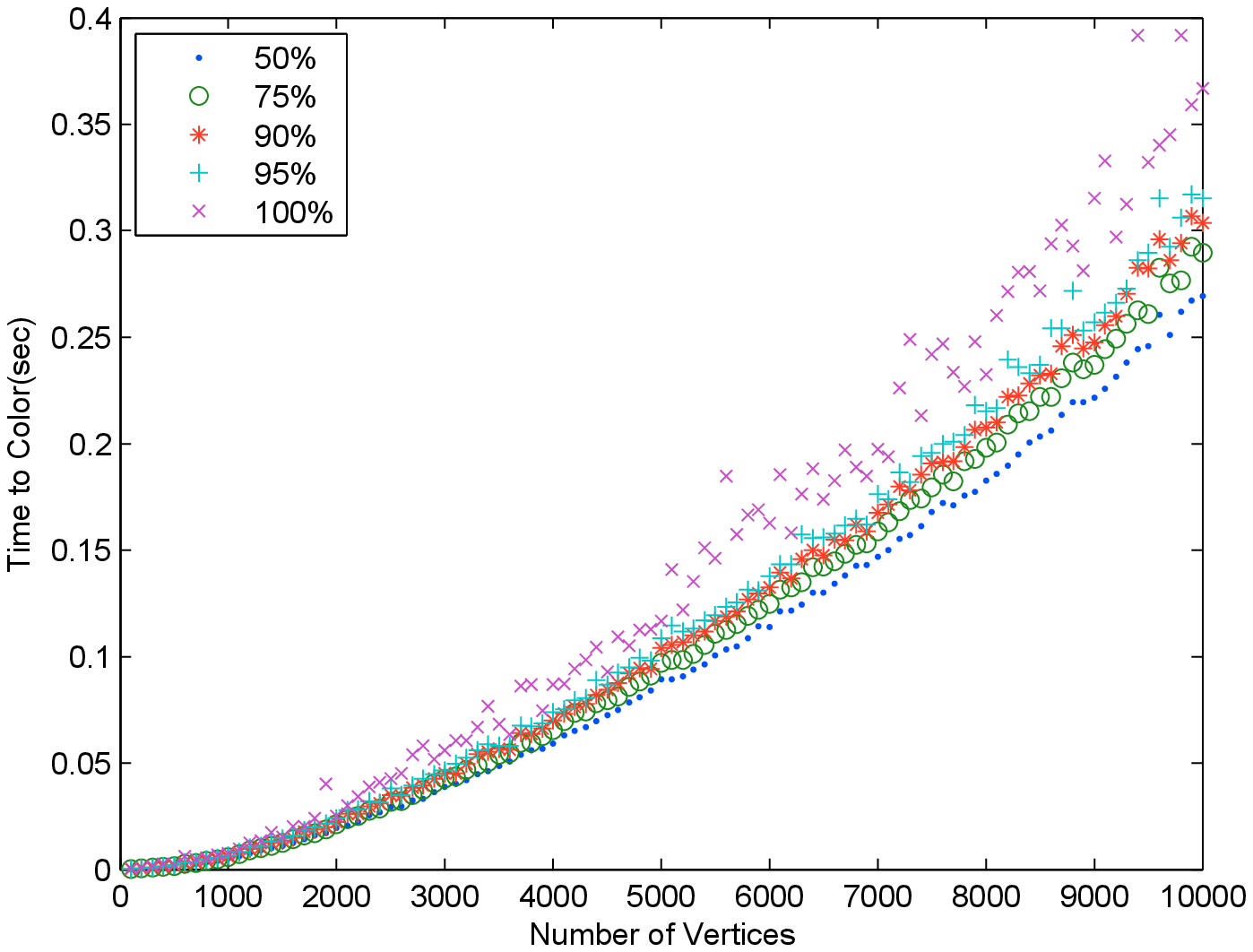}}\\
		\subfloat[Irregular with $\delta=2$]{\includegraphics[width=0.45\textwidth]{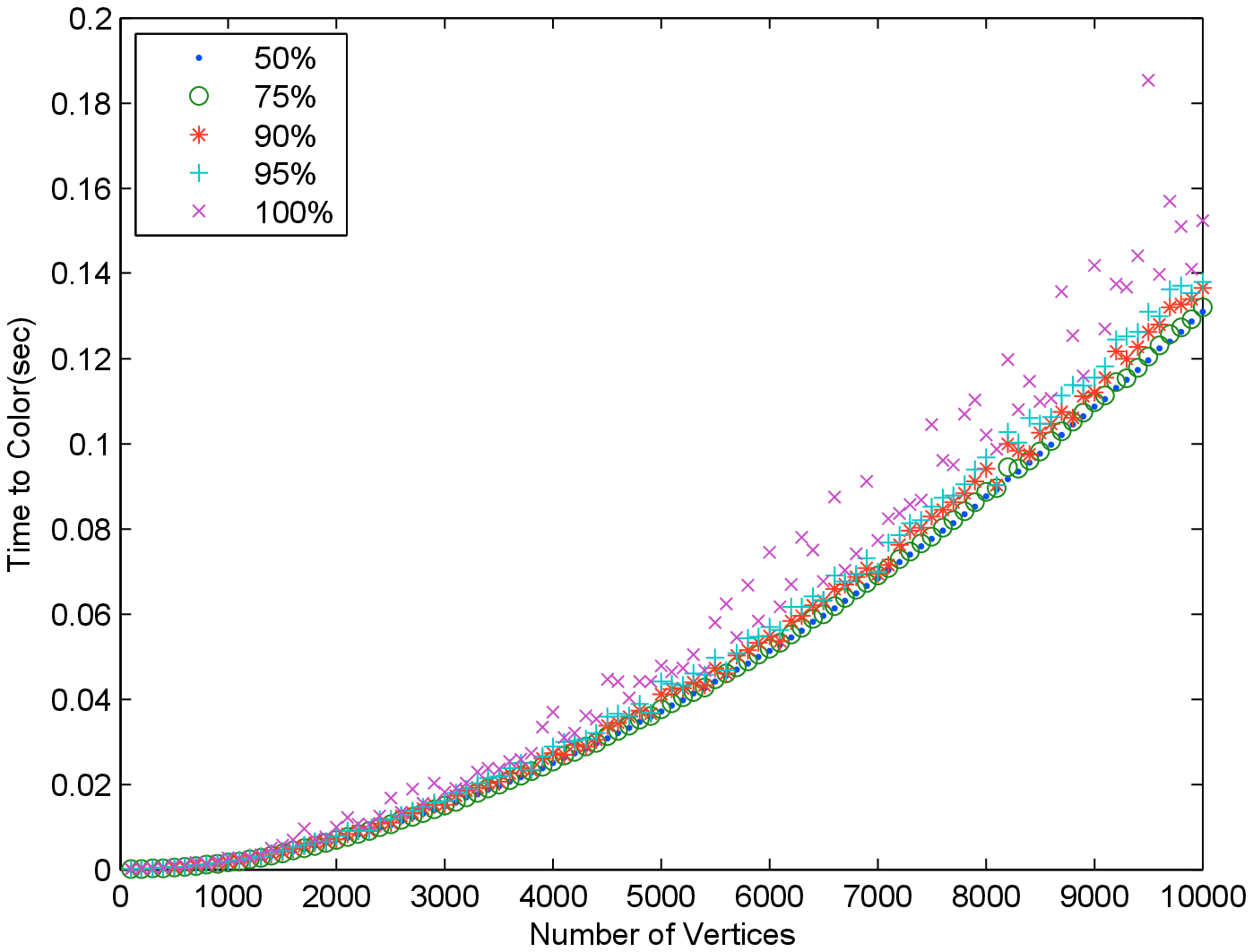}}
		\subfloat[Irregular with $\delta=4$]{\includegraphics[width=0.45\textwidth]{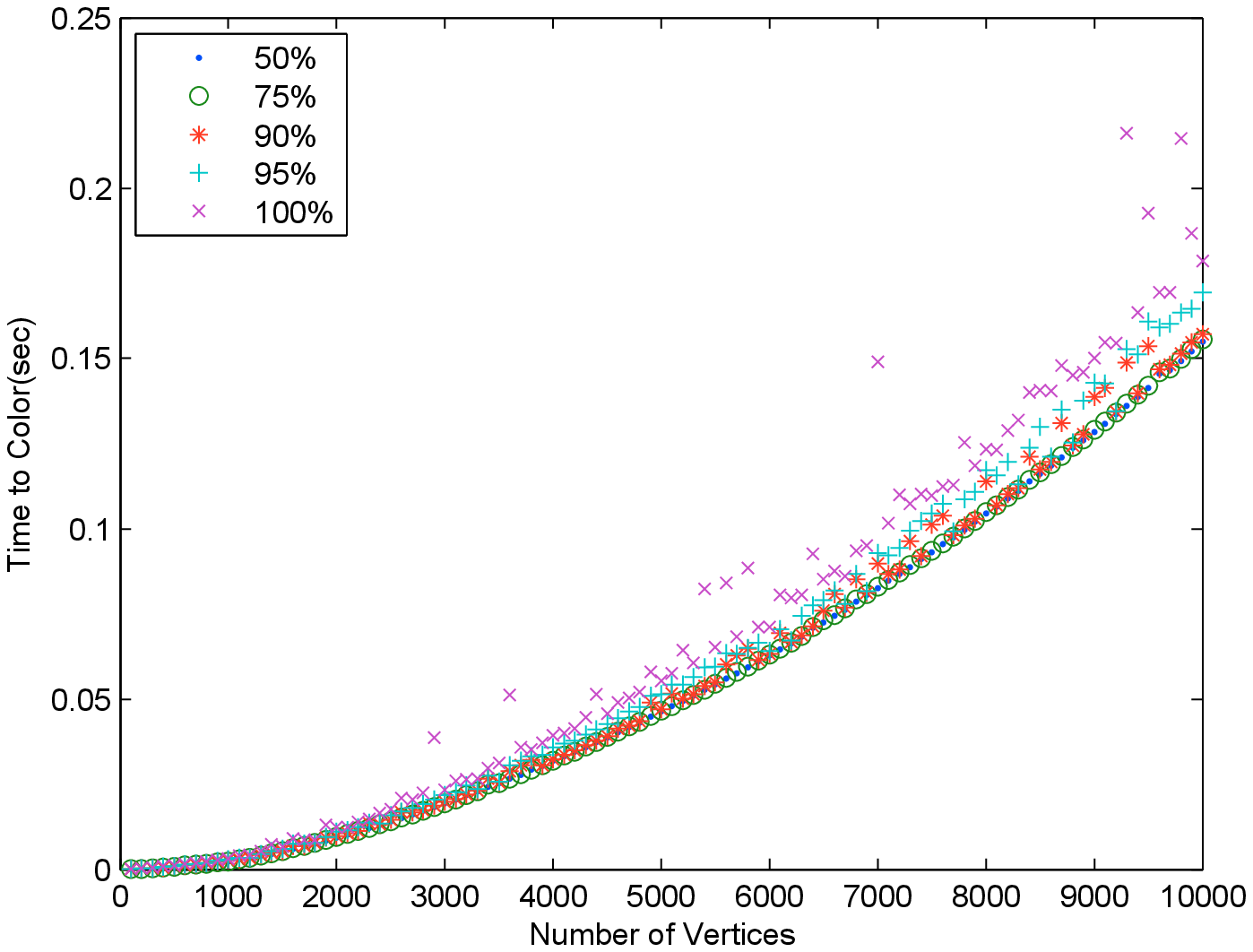}}
		\caption{Experimental running time of edge coloring for random graphs.}
		\label{fig:random}
	\end{figure}

All input random graphs in the testing of our coloring algorithm were generated by the graph generator provided by NetworkX\footnote{NetworkX is a Python package for the creation and manipulation of complex networks, it includes various graph generators as well.}. For regular graphs, 100 instances were randomly generated for each pair of $(\Delta, n)$, where degree $\Delta=4,8$ and the number of vertices $n=100,200,\ldots,10000$. For irregular graphs, 100 instances were randomly generated for each pair of $(\delta,n)$ where density $\delta=|E|/|V|=2,4$ and $n=100,200,\ldots,10000$. The distributions of the running time to color the edges are plotted in Figure~\ref{fig:random}. All randomly generated graphs can be properly $\Delta$-edge-colored by our algorithm. As shown in these figures, the running time of about 95\% of input graphs is highly predictable and quite stable. The experimental results also show that irregular graphs are generally easier to be colored than regular graphs, mainly because of the flexibility of ``coloring'' the don't care edges.
	
\subsection{Results of Benchmark Graphs}

For the purpose of comparisons, we use the same benchmark graphs provided in \cite{hilgemeier2003fast}, which were originally taken from CP2002 \cite{Johnson2002colorinstances}. Since all smaller graphs presented in \cite{hilgemeier2003fast} can be easily solved in a short time by heuristics and our algorithm, we only compare the results of benchmark graphs with more than 500 vertices, which are considered to be large graphs. The experiment results are listed in Table \ref{tab:benchmark_graphs} with graph qualities of each benchmark graph. The best result of each benchmark graph achieved by heuristic algorithms in \cite{hilgemeier2003fast} is compared with the performance of our algorithm. All eleven benchmark graphs considered are $\Delta$-edge-colorable. They were all properly $\Delta$-edge-colored by using our algorithm while only five of them were $\Delta$-edge-colored by the heuristic and evolutionary algorithms (EAs) described in \cite{hilgemeier2003fast}. In Table \ref{tab:benchmark_graphs}, the number of colors used for coloring a benchmark graph $G$ is in {\bf bold} if it is equal to the maximum degree.

Since the performance of these algorithms depends on the particular implementation and computing environment, the absolute time listed in Table \ref{tab:benchmark_graphs} is only for reference, and does not indicate the time complexity of the algorithm. Nevertheless, these experiment results clearly reveal that our algorithm outperforms heuristic algorithms in accuracy and efficiency.

	\begin{table}[thbp]
		\centering
		\footnotesize
		\begin{tabular}{|c||c|c|c|c|c||c|c|c|c|}
			\hline
			\multicolumn{6}{|c||}{Graph Qualities} & \multicolumn{2}{c|}{Best Heuristic in \cite{hilgemeier2003fast}} & \multicolumn{2}{c|}{Our Algorithms} \\
			name & vertices & edges & $\delta$ & $\Delta$ & $\chi_e(G)$ & colors & secs & colors & secs \\
			\hline\hline
			DJSC500.1 & 500 & 12458 & 24.92 & 68 & \emph{68} & 69 & 0.25 & \textbf{68} & 0.0227 \\
			ash331GPIA & 662 & 4185 & 6.32 & 23 & 23  & \textbf{23} & 0.01 & \textbf{23} & 0.0106 \\
			ash958GPIA & 1916 & 12506 & 6.53 & 24 & 24 & \textbf{24} & 0.01 & \textbf{24} & 0.0766 \\
			will199GPIA & 701 & 6772 & 9.66 & 38 & \emph{38} & 40 & 0.03 & \textbf{38} & 0.0158 \\
			4-FullIns\_4 & 690 & 6650 & 9.64 & 119 & 119 & 120 & 0.03 & \textbf{119} & 0.0114 \\
			5-FullIns\_4 & 1085 & 11395 & 10.50 & 160 & 160 & 161 & 0.09 & \textbf{160} & 0.0264 \\
			qg.order30 & 900 & 26100 & 29.00 & 58 & 58 & \textbf{58} & 0.59 & \textbf{58} & 0.2174 \\
			qg.order60 & 3600 & 212400 & 59.00 & 118 & 118 & \textbf{118} & 17.94 & \textbf{118} & 7.4265 \\
			qg.order100 & 10000 & 990000 & 99.00 & 198 & \emph{198} & 212 & 248.38 & \textbf{198} & 94.0123 \\
			wap04a & 5231 & 294902 & 56.38 & 351 & 351 & \textbf{351} & 24.98 & \textbf{351} & 3.2052 \\
			latin\_square\_10 & 900 & 202081 & 224.53 & 512 & \emph{512} & 554 & 268.85 & \textbf{512} & 2.1981 \\
			\hline
		\end{tabular}
		\caption{Performance on benchmark graphs.}
		\label{tab:benchmark_graphs}
	\end{table}

\section{Conclusions}

In this paper, edge coloring of simple graphs is solved by a variable elimination process similar to the solving of linear equations. The connections between graphs and linear equations provide cornerstones in many areas such as electric circuit theory and Markov chains. In edge coloring of simple graphs, variables are eliminated by color-exchange operations implemented on graphs. The problem is solved by a sequence of configuration transformations in the same manner as solving the puzzle of Rubik's Cube, which has a final configuration that can always be reached from any initial configuration. In the case of edge coloring of graphs, however, only $\Delta$-edge-colorable graphs have final configurations.

Another related problem that could be solved by color exchanges is finding the Hamiltonian cycles. A simple graph $G$ may have more than one proper color configurations. Consider the set of all proper color configurations as the state space of a Markov chain associate with edge-colored graph $G$. A state is Hamiltonian if it contains a two-colored Hamiltonian cycle, which can be reached by random walks on the Markov chain. In the future, the application of the algebraic method proposed in this paper to graph factors and Hamiltonian cycles could be challenging research topics.

\bibliography{reference} % reference.bib
\bibliographystyle{abbrv}

\section*{Appendix}

\begin{figure}[htpb]
	\centering
	\includegraphics[scale=0.85]{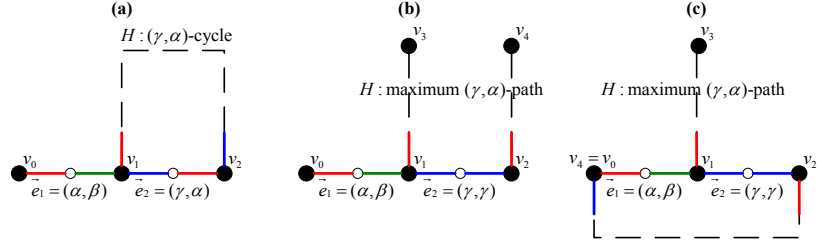}
	\caption{Illustration of non-Kempe walk {\bf Case DW1}.}
	\label{fig:nonKempe1}
\end{figure}

An effective one-step non-Kempe move of the tagged variable $\vec{e}_1=(\alpha,\beta)$  to $\vec{e}_2=(\gamma,\delta)$, where $\gamma \neq \alpha$ and $\delta \neq \beta$, requires one of the following two types of color inversion on the next edge $\vec{e}_2$: 
\begin{align}
\alpha \mbox{-type}: \vec{e}_2 & =(\gamma,\delta) \rightarrow \vec{e}_2=(\alpha,*) \nonumber \\ 
\beta \mbox{-type}: \vec{e}_2 & =(\gamma,\delta) \rightarrow \vec{e}_2=(*,\beta) \nonumber 
\end{align}
A color inversion applied to $\vec{e}_2$ may become invalid if the operation also involves $\vec{e}_1$. The following two cases are considered in a canonical configuration:

{\bf Case DW1}: $\vec{e}_1 \circ \vec{e}_2 =(\alpha,\beta) \circ (\gamma,\delta)$ , where $\gamma \neq \alpha , \delta \neq \beta$ and $\delta \in \{\alpha, \gamma \}$.
\begin{itemize}  
\itemsep -2pt %reduce space between items
\item {\bf (1.1)} $\delta=\alpha$, $\alpha$-type: inversion of the $(\gamma,\alpha)$ cycle $H$ shown in Fig. \ref{fig:nonKempe1} (a) that contains $\vec{e}_2=(\gamma,\alpha)$
\item {\bf (1.2)} $\delta = \gamma$, $\alpha$-type: inverse the maximum $(\gamma,\alpha)$ path $H$ shown in Fig. \ref{fig:nonKempe1} (b) that contains the edge $\vec{e}_2=(\gamma,\gamma)$.\\
The variable $\vec{e}_1=(\alpha,\beta)$ is {\it blocked} if $v_3=v_0$. Note that the variable $\vec{e}_1$  is not blocked if $v_4=v_0$, as illustrated in Fig. \ref{fig:nonKempe1} (c), the following sequence of operations can move $\vec{e}_1$ one step forward:
\begin{enumerate}
\itemsep -2pt %reduce space between items
\item exchange color on the interior chain from $v_2$ to $v_0$, hence $\vec{e}_1=(\gamma,\beta)$ and $\vec{e}_2=(\gamma,\alpha)$.
\item $\vec{e}_1 \otimes \vec{e}_2 = (\gamma,\beta) \otimes (\gamma,\alpha) \Rightarrow (\gamma,\gamma) \circ (\beta,\alpha)$
\end{enumerate}
\end{itemize}

\begin{figure}[htpb]
	\centering
	\includegraphics[scale=0.85]{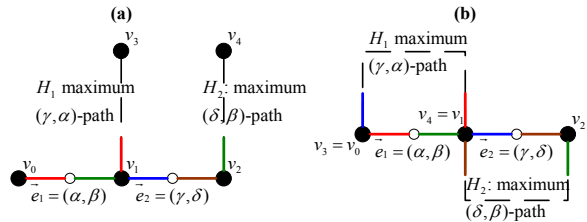}
	\caption{Illustration of non-Kempe walk {\bf Case DW2}.}
	\label{fig:nonKempe2}
\end{figure}

{\bf Case DW2}: $\vec{e}_1 \circ \vec{e}_2 =(\alpha,\beta) \circ (\gamma,\delta)$ , where $\gamma \neq \alpha , \delta \neq \beta$ and $\delta \notin \{\alpha, \gamma \}$.
\begin{itemize}  
\itemsep -2pt 
\item {\bf (2.1)} $\alpha$-type: inverse the maximum $(\gamma,\alpha)$ path $H_1$ shown in Fig. \ref{fig:nonKempe2} (a) that contains the vertex $v_1$, or
\item {\bf (2.2)} $\beta$-type: inverse the maximum $(\delta,\beta)$ path $H_2$ shown in Fig. \ref{fig:nonKempe2} (a) that contains the vertex $v_2$.
\end{itemize}

The variable $\vec{e}_1=(\alpha,\beta)$  is {\it blocked} if $v_3=v_0$  and $v_4=v_1$, as illustrated in Fig. \ref{fig:nonKempe2} (b).

Table \ref{tab:nonKempeWalkcase} lists all possible one-step non-Kempe moves of the tagged variable $\vec{e}_1=(\alpha,\beta)$.

\begin{table}[!t]
	\centering
	{\footnotesize
		\begin{tabular}{|c|c| c| c|c| }
			\hline 
			Case & Next Step & Color Inversion& Operation & Result \\
			\hline \hline
			DW1.1 & $(\alpha,\beta)\circ (\gamma,\alpha)$ & $(\gamma,\alpha) \rightarrow (\alpha,\gamma) $ & $(\alpha,\beta)\otimes(\alpha,\gamma)\Rightarrow(\alpha,\alpha)\circ(\beta,\gamma)$&eliminate one variable\\ \hline	
			DW1.2 & $(\alpha,\beta)\circ (\gamma,\gamma)$ & $(\gamma,\gamma) \rightarrow (\alpha,\alpha)$ & $(\alpha,\beta)\otimes(\alpha,\alpha) \Rightarrow (\alpha,\alpha)\circ(\beta,\alpha)$& step forward \\ \hline 
			blocked DW1.2    &                                        &                                              & if $v_3=v_0$  & blocked \\ \hline 
			DW2.1 & $(\alpha,\beta)\circ (\gamma,\delta)$  & $(\gamma,\delta) \rightarrow (\alpha,\delta)$ & $(\alpha,\beta)\otimes(\alpha,\delta)\Rightarrow (\alpha,\alpha)\circ(\beta,\delta)$& eliminate one variable \\ \hline 
			DW2.2 & $(\alpha,\beta)\circ (\gamma,\delta)$ & $(\gamma,\delta) \rightarrow (\gamma,\beta)$ &$(\alpha,\beta)\otimes(\gamma,\beta)\Rightarrow(\alpha,\gamma) \circ (\beta,\beta)$& eliminate one variable \\ \hline	
		 blocked DW2    &                                       &                                              & if $v_3=v_0$  and $v_4=v_1$ & blocked \\ \hline	
		\end{tabular}
		}
	\caption{One-step non-Kempe moves of $(\alpha,\beta)$ variable on directional path.}
	\label{tab:nonKempeWalkcase}
\end{table}

\end{document}